\documentclass[%
 reprint,
preprintnumbers,
 amsmath,amssymb,
 aps,
]{revtex4-2}

\usepackage{graphicx}
\usepackage{dcolumn}
\usepackage{bm}

\begin{document}

\preprint{FERMILAB-PUB-23-495-T}

\title{Beyond Generalized Eigenvalues in Lattice Quantum Field Theory}

\author{George T. Fleming}
\email{gfleming@fnal.gov}
\affiliation{%
Theory Division, Fermi National Accelerator Laboratory, Batavia, IL
}%

\date{\today}

\begin{abstract}
Two analysis techniques, the \textit{generalized eigenvalue method} (GEM) or
\textit{Prony's (or related) method} (PM), are commonly used to analyze statistical
estimates of correlation functions produced in lattice quantum field theory calculations.
GEM takes full advantage of the matrix structure of correlation functions but only considers
individual pairs of time separations when much more data exists.  PM can be applied to many
time separations and many individual matrix elements simultaneously but does not fully
exploit the matrix structure of the correlation function. We combine both these methods
into a single framework based on \textit{matrix polynomials}.  As these algebraic methods
are well known for producing extensive spectral information about statistically-noisy data,
the method should be paired with some information criteria, like the recently proposed
Bayesean model averaging.
\end{abstract}

\maketitle


\section{\label{sec:intro}Introduction}

We briefly review the construction of lattice correlation functions, which has been the
subject of many reviews, \textit{e.g.}\ \cite{Cohen:2010}, and even textbooks. Our review
is not comprehensive and serves only to fix the notation.

A standard computational task of lattice quantum field theory is to generate statistical
samples of Euclidean time correlation functions, which are often written as vacuum-to-vacuum
transition amplitudes
\begin{eqnarray}
\lefteqn{C_{ab}(\vec{p},|t-t_0|) =} \nonumber \\*
& & \sum_{\vec{x}} e^{i \vec{p} \cdot ( \vec{x} - \vec{x}_0 )}
\left\langle 0 \left|
  \mathcal{O}_a(\vec{x},t) \mathcal{O}_b^\dagger(\vec{x}_0,t_0)
\right| 0 \right\rangle
\end{eqnarray}
where we have made explicit use of translation invariance of the ensemble average
so that it is clear the correlation function is independent of the spatial source location
$\vec{x}_0$ and only depends on the relative time separation of the source and sink
$|t - t_0|$. $\mathcal{O}_b^\dagger(\vec{x}_0,t_0)$ is a creation operator that creates a
quantum state with a given set of quantum numbers associated with a particular irreducible
representation of the lattice Hamiltonian and $\mathcal{O}_a(\vec{x},t)$ is an annihilation
operator that annihilates a quantum state of a given set of quantum numbers.  If the two
sets of quantum numbers are not commensurate then the correlation is zero by symmetry.
There are many possible operators that can create and annihilate a given set of quantum
numbers, so we give them labels $a, b \in [1,N]$ and the correlation function can be
a $\mathbb{C}^{N \times N}$ matrix in this operator space.

A standard exercise is to expand this correlation function in a complete set of eigenstates
of the lattice Hamiltonian.  Associated with each such state is an energy
eigenvalue $E_k$ and eigenvector $\left| k \right\rangle$.
For simplicity, we have suppressed labels associated
with other conserved quantum numbers.  The correlation function is now written
\begin{equation}
\label{eq:corr}
C_{ab}(\vec{p},|t-t_0|) =
\sum_k z_{ak}(\vec{p}) \ z_{bk}^*(\vec{p}) \ e^{-E_k(\vec{p})\ t}
\end{equation}
but from here on we will suppress the dependence on momentum $\vec{p}$ and freely
assume time translation invariance. We will assume
for each creation operator $\mathcal{O}_a^\dagger$
a conjugate annihilation operator
can be defined $\mathcal{O}_a$ such that
$z_{ak} = \left\langle 0 \left| \mathcal{O}_a \right| k \right\rangle$ and
$z_{ak}^* = \left\langle k \left| \mathcal{O}_a^\dagger \right| 0 \right\rangle$.
We will also assume that any operators that are not linearly independent have
been removed from the operator basis and so $z_{ak}$ can be viewed as elements
of a matrix with full rank. In matrix notation, the correlation function can
be written
$\bm{C}(t) = \bm{Z} \bm{\Lambda}^t \bm{Z}^\dagger$.
Formally, the total number of eigenstates should be considered very much larger
than the dimension of $\bm{C}(t)$ that can be computed
in a standard lattice calculation.  
Hence, the typical data analysis problem is to extract as much \textit{reliable}
information as possible about the eigenvalue spectrum $E_k$ and the matrix elements
$z_{ak}$ given a finite number of samples of the matrix function $\bm{C}(t)$.

Throughout the paper we use the following notation:
\begin{description}
\item[$C_{ab}(t)$] Scalar correlation function $\in \mathbb{C}^{1 \times 1}$.
\item[$t$] Integer-spaced Euclidean time separations.
\item[$\bm{C}(t)$, $\bm{C}_t$] Matrix correlation function
$\in \mathbb{C}^{N \times N}$.
\item[$N$] Dimension of operator basis.
\item[$a, b, n$] Operator indices $\in [1, N]$.
\item[$L$] Exact polynomial order (could be $\infty$).
\item[$K$] Assumed polynomial order.
\item[$\ell$] Index $\in [0, L-1]$ or $[0,K-1]$.
\item[$k$] Index of states contributing to $\bm{C}(t)$.
\item[$\lambda_k$] $\exp(-E_k)$ energy contributing to $\bm{C}(t)$.
\item[$\bm{\Lambda}$] Diagonal matrix of energies
  $\in \mathbb{R}^{L N \times L N}$ or $\mathbb{R}^{K N \times K N}$.
\item[$z_{ak}$] $\left\langle 0 \left| \mathcal{O}_a \right| k \right\rangle$
 amplitude contributing to $\bm{C}(t)$.
\item[$\bm{Z}$] Matrix of amplitudes $\in \mathbb{C}^{N \times L N}$ or
  $\mathbb{C}^{N \times K N}$
\item[$\bm{\mathcal{C}}$] Companion matrix $\in \mathbb{C}^{L N \times L N}$
  or $\mathbb{C}^{K N \times K N}$.
\item[$\bm{P}_\ell$] Linear prediction matrices
$\in \mathbb{C}^{N \times N}$.
\item[$p(M|D)$] Bayesian model probability \cite{Jay:2020jkz}.
\item[$q$] Number of model parameters, usually $L N (N+1)$ or $K N (N+1)$.
\end{description}

\subsection{\label{sub:GEM}Generalized Eigenvalue Method}

Early in the development of lattice quantum field theory, Wilson recognized
\cite{Wilson:1981, Berg:1982hf}
that correlation function data could be analyzed by truncating the expansion
in Eq.~(\ref{eq:corr}) to a relatively small
number of states and that the energies computed in such a fashion
would be variational estimates.
A simple method to analyze a scalar correlation function is called
\textit{effective energy}
\begin{equation}
\label{eq:eff_energy}
\lambda_{\text{eff}}(t,t_0)^{t-t_0}
= e^{-E_{\text{eff}}(t,t_0) \ (t - t_0)} = \frac{C(t)}{C(t_0)}
\end{equation}
Asymptotically, $E_\text{eff}(t,t_0) = E_1 + \mathcal{O}(e^{-(E_2-E_1) t})$
as $t \to \infty$, where $E_1, E_2$ are the lowest and next-to-lowest
energy eigenvalues, assuming the states created by the
operators used in the construction have at least some overlap with the true
ground state,
$\left\langle 0 \left| \mathcal{O}_a \right| 1 \right\rangle \ne 0$.

The method of using generalized eigenvalues to extract information
about the spectrum of the Hamiltonian is well-studied in lattice QFT
\cite{Michael:1982gb, Luscher:1990ck, Blossier:2009kd}.
For simplicity of discussion, we will assume that only $N$ states
contribute to the Hamiltonian and that we have a complete basis of $N$ operators
so the matrix $\bm{Z}$ is of full rank and invertable.  Generalized eigenvalues
are computed by solving the following problem
\begin{equation}
\label{eq:GE_eqn}
\lambda_{\text{eff},k}(t,t_0)^{t-t_0} \ \bm{C}(t_0) \ \bm{v}_k(t,t_0)
= \bm{C}(t) \ \bm{v}_k(t,t_0)
\end{equation}
In our simple case, if $\bm{v}_k$ is the $k$-th column vector
of $[ \bm{Z}^\dagger ]^{-1}$ then $\lambda(t,t_0) = e^{-E_k}$ and there
will be $N$ such solutions.  Since the general case is that the number
of Hamiltonian eigenstates is much larger than the dimension
of the operator basis,
the generalized eigenvalues $\lambda(t,t_0)$ will only be variational estimates
and will depend on the choice $(t,t_0)$.  Thus, the generalized eigenvalue method
can be viewed as an extension of the effective energy method to matrix-valued
correlation functions.

One approach to bounding the generalized eigenvalues is the \textit{eigenvalue
interlacing theorem}.  Simply stated, if $\bm{A} \in \mathbb{C}^{K \times K}$
is a Hermitian matrix with eigenvalues
$\alpha_1 \ge \alpha_2 \ge \cdots \ge \alpha_K$ and
$\bm{B} \in \mathbb{C}^{N \times N}, N < K$ is also a Hermitian matrix
with eigenvalues $\beta_1 \ge \cdots \ge \beta_N$ and an orthogonal projection
matrix $\bm{P} \in \mathbb{C}^{N \times K}$ exists such that
$\bm{P} \bm{A} \bm{P}^\dagger = \bm{B}$ then the eigenvalues interlace:
$\alpha_k \ge \beta_k \ge \alpha_{k+K-N}$. To apply these bounds to generalized
eigenvalues requires $\bm{C}(t_0)$ to be invertable so the problem can be
transformed to a standard eigenvalue problem.  Then,
$\lambda_k \ge \lambda_{\text{eff},k}(t,t_0) \ge \lambda_{k+K-N}$ which is true
for any $(t,t_0)$. Of course, if $K \gg N$ then the right hand bound
is not very restrictive but the more important bound is to the left which indicates
the generalized eigenvalues are always an over-estimate of the true eigenvalue.

Asymptotic bounds for $\lambda_{\text{eff},k}(t,t_0)$ as $t \to \infty$
are also known \cite{Blossier:2009kd}.  For fixed $t_0$ as $t \to \infty$
\cite{Luscher:1990ck}
\begin{eqnarray}
E_{\text{eff},k}(t,t_0) & = &  E_k + \mathcal{O}\left(e^{-\Delta E_k t}\right)
\nonumber \\*
\Delta E_k & = & \min_{m \ne k} \left| E_m - E_k \right| .
\end{eqnarray}
If $t_0$ is increased such that $t/2 \le t_0$ as $t \to \infty$ then
\begin{equation}
\label{eq:eff_energy_error}
E_{\text{eff},k}(t,t_0) = E_k +
\mathcal{O}\left(e^{-(E_{N+1} - E_k) t}\right) .
\end{equation}

It would be desirable to find a method that extends the generalized eigenvalue method
to simultaneously use the values of matrix-valued correlation functions
computed for more than two
time separations.  The approach we would like to consider is viewing the generalized
eigenvalue method as a matrix polynomial \cite{Gohberg:2009} of linear degree and ask
whether we can construct higher degree matrix polynomials using more time separations.
An monic $N \times N$ matrix polynomial of degree $L$ is
\begin{equation}
\bm{\mathcal{P}}_L(\lambda) = \bm{I} \lambda^L + \sum_{\ell=0}^{L-1} \bm{P}_\ell \lambda^\ell,
\ \bm{P}_\ell \in \mathbb{C}^{N \times N}
\end{equation}
Generally, $\det\bm{\mathcal{P}}_L(\lambda_k) = 0$ admits
$N L$ solutions $\lambda_k$.
If the matrix coefficient of $\lambda^L$ is not the identity, as long as it's invertable,
then the monic form can be constructed.  For example, a first-order monic matrix
polynomial can be derived from Eq.~(\ref{eq:GE_eqn})
\begin{equation}
\label{eq:matrix_poly}
\bm{\mathcal{P}}_1(\lambda) = \bm{I} \lambda^{t-t_0}
- {\bm{Q}^\dagger}^{-1} \ \bm{C}(t) \ \bm{Q}^{-1}
\end{equation}
where $\bm{C}(t_0) = \bm{Q}^\dagger \bm{Q}$
and $\det\bm{\mathcal{P}}_1(\lambda_k) = 0$ has the same $N$ solutions
as Eq.~(\ref{eq:GE_eqn}).

However, it is not sufficient to construct
just any matrix polynomial.  For example, we could add Eq.~(\ref{eq:GE_eqn}) twice
to form
$2 \bm{C}(t_0+2) \bm{v} = \lambda \bm{C}(t_0 + 1) \bm{v} + \lambda^2 \bm{C}(t_0) \bm{v}$,
which would be equivalent to a second-order matrix polynomial.
While this equation yields two solutions in the $N=1$ case, we leave as an exercise
for the reader to see that the solutions are not exact when only two states contribute
to the correlation function \cite{Fleming:2004hs}. So, our goal is to construct degree $L$
matrix polynomials from the correlation function data that admit $N L$ solutions and those
solutions are the exact solutions when only $N L$ states contribute.  In particular,
the matrix polynomials should be equivalent to the polynomials produced by Prony's method
when $N=1$, since those are known to produce exact solutions.

\subsection{\label{sub:PM}Prony's (and related) Method}

Another extension of the effective energy method would be Prony's method, first
discovered in 1795 \cite{Prony:1795}, and since rediscovered many times
in several related guises.
The method extracts multiple effective energies by using a single correlation
function at more than two time separations.  The basic idea is to assume
that the correlation function over a range of time separations can be well-modeled
by a spectrum of $L$ effective energies $\lambda_k = \exp(-E_k)$.  This spectrum
is used to construct a characteristic polynomial
\begin{equation}
\mathcal{P}_L(\lambda) = \prod_{k=1}^L (\lambda - \lambda_k) = \lambda^L
+ \sum_{\ell=0}^{L-1} p_\ell \lambda^{\ell}
\end{equation}
whose \textit{linear prediction} coefficients $p_\ell$ can be computed as functions
of the input spectrum $\lambda_k$. $\mathcal{P}_L(\lambda_k) = 0$ can be solved
\begin{equation}
\label{eq:linear_pred}
\lambda_k^t = \lambda_k^{t-L} \lambda_k^{L}
= - \lambda_k^{t-L} \sum_{\ell=0}^{L-1} p_\ell \lambda_k^{\ell},
\ t \ge L
\end{equation}
and used in combination with Eq.~(\ref{eq:corr}) for a scalar function
with a compact notation, $C_t = C(t)$ to write the linear system
\begin{equation}
\label{eq:linear_prediction_hankel_system}
\left[ \begin{array}{c}
  C_L \\ C_{L+1} \\ \vdots \\ C_{2 L - 1}
\end{array} \right] = - \left[ \begin{array}{ccc}
  C_0     & \cdots & C_{L-1} \\
  C_1     & \cdots & C_{L}   \\
  \vdots  & \ddots & \vdots  \\
  C_{L-1} & \cdots & C_{2L-2}
\end{array} \right] \left[ \begin{array}{c}
  p_0 \\ p_1 \\ \vdots \\ p_{L-1}
\end{array} \right]
\end{equation}
The structured matrix shown here is commonly called a Hankel matrix even though
Hankel's work in this area occurred nearly a century later.
Prony's solution is to take the computed scalar correlation function data
$C_t$ and solve the linear system to find the unknown $p_\ell$.  Then, given
the $p_\ell$, the polynomial equation $\mathcal{P}_L(\lambda) = 0$ can be solved to find
the $\lambda_k$.

Once the solution for the $\lambda_k$ has been computed, the coefficients
can be determined by solving another structured linear system
\begin{equation}
\label{eq:Vandermonde}
\left[\begin{array}{c}
  C_0     \\
  C_1     \\
  C_2     \\
  \vdots  \\
  C_{2 L-1} \\
\end{array}\right] = \left[\begin{array}{cccc}
  1          & 1          & \cdots & 1          \\
\lambda_1   & \lambda_2   & \cdots & \lambda_L   \\
\lambda_1^2 & \lambda_2^2 & \cdots & \lambda_L^2 \\
\vdots      & \vdots      & \ddots & \vdots      \\
\lambda_1^{2 L-1} & \lambda_2^{2 L-1} & \cdots & \lambda_L^{2 L-1} \\
\end{array}\right] \left[\begin{array}{c}
  a_1    \\
  a_2    \\
  \vdots \\
  a_L
\end{array}\right]
\end{equation}
using singular value decomposition (SVD).
The structured matrix here is called a Vandermonde matrix. Vandermonde, at least,
was a contemporary of Prony but it's not clear the connection between their work
was recognized at the time.  There have been numerous approaches to combining data
from multiple matrix elements but we have yet to see an approach that fully exploits
the structure of the matrices.  In particular, as the scalar Prony method involves
solving a polynomial equation to find the spectrum, we anticipate that a fully
structured approach to the matrix problem will involve solving a matrix polynomial.

In preparation for constructing a block Prony method, we see how a naive application
of linear prediction applied element-by-element
to correlator matrices fails in a case where only two states contribute
to the correlation function
\begin{equation}
C_{ab}(t) = \lambda_1^t z_{a1} z_{b1}^* + \lambda_2^t z_{a2} z_{b2}^*
\end{equation}
Now, use $L=2$ linear prediction
\begin{eqnarray}
C_{ab}(2) & = & - ( p_1 \lambda_1 + p_0 ) z_{a1} z_{b1}^*
  -(p_1 \lambda_2 + p_0 ) z_{a2} z_{b2}^* \nonumber \\*
& = & - p_1 C_{ab}(1) - p_0 C_{ab}(0) \\
\label{eq:naive_block_linear_prediction}
\bm{C}_t & = & - \bm{C}_{t-1} ( p_1 \bm{I} ) - \bm{C}_{t-2} ( p_0 \bm{I} )
\end{eqnarray}
Substituting Eq.~(\ref{eq:naive_block_linear_prediction}) into
Eq.~(\ref{eq:linear_prediction_hankel_system}) reveals the problem:
solving the block linear system does not guarantee the resulting linear prediction
matrices $\bm{P}_0$, $\bm{P}_1$ are proportional to the identity matrix
or even have degenerate eigenvalues.

\section{Block Prony Method}

As previously mentioned, there are a number of previous attempts to incorporate
matrix-valued correlation functions into Prony's method, or other
algebraically-related methods \cite{Fleming:2009wb, Beane:2009kya,
Cushman:2019hfh, Fischer:2020bgv}.
We will name this method the \textit{Block Prony Method} (BPM)
as we are not aware of any other method called by this name.
To demonstrate the construction, we will assume
we are given $2 L$ Hermitian positive definite correlation matrices
$\bm{C}_t \in \mathbb{C}^{N \times N}$ computed at equally-spaced time
separations $t \in [0, 2 L - 1]$.  We will also assume that precisely
$N L$ energies
contribute to the correlation matrices, $\lambda_k = \exp(-E_k)$
and $k \in [1, N L]$, according to Eq.~(\ref{eq:corr}).  A matrix
$\bm{\mathcal{C}} \in \mathbb{C}^{N L \times N L}$, called the second companion
matrix, yields the correct characteristic polynomial
$\det [ \lambda \bm{I} - \bm{\mathcal{C}} ]
\propto \prod_{k=1}^{N L} (\lambda - \lambda_k)$.  In terms of
$\mathbb{C}^{N \times N}$ blocks
\begin{equation}
\bm{\mathcal{C}} = \left[
\begin{array}{ccccc}
0      & 0      & \cdots & 0      & - \bm{P}_0     \\
\bm{I} & 0      & \cdots & 0      & - \bm{P}_1     \\
0      & \bm{I} & \ddots & \vdots & - \bm{P}_2     \\
\vdots & \ddots & \ddots & 0      & \vdots         \\
0      & \cdots & 0      & \bm{I} & - \bm{P}_{L-1}
\end{array}
\right]
\end{equation}
and the matrix polynomial $\bm{\mathcal{P}}_L(\lambda)$
of Eq.~(\ref{eq:matrix_poly}) using these blocks satisfies
the same characteristic equation. To derive the block equivalent
of Eq.~(\ref{eq:linear_pred}), we consider an eigenvector
$\bm{v}_k$ of $\bm{\mathcal{C}}$ partitioned into
$L$ $\mathbb{C}^{1 \times N}$ blocks
\begin{equation}
\lambda_k \left[\begin{array}{c}
\bm{v}_{k,0} \\
\bm{v}_{k,1} \\
\vdots \\
\bm{v}_{k,L-1}
\end{array} \right] = \left[
\begin{array}{cccc}
0      & 0      & \cdots & - \bm{P}_0     \\
\bm{I} & \ddots & \cdots & - \bm{P}_1     \\
\vdots & \ddots & 0      & \vdots         \\
0      & \cdots & \bm{I} & - \bm{P}_{L-1}
\end{array}
\right]
\left[\begin{array}{c}
\bm{v}_{k,0} \\
\bm{v}_{k,1} \\
\vdots \\
\bm{v}_{k,L-1}
\end{array} \right]
\end{equation}
which can also be written
\begin{eqnarray}
\lambda_k \bm{v}_{k,0} & = & - \bm{P}_0 \bm{v}_{k,L-1} \\
\lambda_k \bm{v}_{k,\ell} & = & \bm{v}_{k,\ell-1}
  - \bm{P}_{\ell} \bm{v}_{k,L-1} .
\end{eqnarray}
These equations can be reduced by substitution to
\begin{equation}
\lambda_k^L \bm{v}_{k,L-1} = \left(
 - \sum_{\ell=0}^{L-1} \lambda_{k}^{\ell} \bm{P}_\ell
\right) \bm{v}_{k,L-1}
\end{equation}
Given our assumptions, this can be repeated for any linear
combination of eigenvectors of $\bm{\mathcal{C}}$ which enables
us to extend Eq.~(\ref{eq:linear_prediction_hankel_system})
to block form
\begin{equation}
\label{eq:linear_prediction_hankel_system_block}
\left[ \begin{array}{c}
  \bm{C}_L \\ \bm{C}_{L+1} \\ \vdots \\ \bm{C}_{2 L - 1}
\end{array} \right] = - \left[ \begin{array}{ccc}
  \bm{C}_0     & \cdots & \bm{C}_{L-1} \\
  \bm{C}_1     & \cdots & \bm{C}_{L}   \\
  \vdots       & \ddots & \vdots       \\
  \bm{C}_{L-1} & \cdots & \bm{C}_{2L-2}
\end{array} \right] \left[ \begin{array}{c}
  \bm{P}_0 \\ \bm{P}_1 \\ \vdots \\ \bm{P}_{L-1}
\end{array} \right]
\end{equation}
Now, the algorithm is clear.  Given the $\bm{C}_t$, solve
Eq.~(\ref{eq:linear_prediction_hankel_system_block})
to find the $\bm{P}_\ell$ and then construct the second companion
matrix $\bm{\mathcal{C}}_L$ and solve for the eigenvalues
$\lambda_k$.  As an aside, if more than $2 L$ equally-spaced $\bm{C}_t$
are available, additional rows could be appended to the block Hankel in
Eq.~(\ref{eq:linear_prediction_hankel_system_block}) while keeping the number
of columns fixed.  This transforms the problem into an overconstrained system
could be analyzed by least squares using SVD.

\section{Bilinear System (BLS) of Equations}

Once the $\lambda_k$ have been computed following the block Prony method
of the previous section,  it is desirable to use the given $\bm{C}_t$
and the previously computed $\lambda_t$ to compute the coefficients
$z_{a k}$ of Eq.~(\ref{eq:corr}).  In the scalar case, this can be done by solving
the Vandermonde system Eq.~(\ref{eq:Vandermonde}) but in the matrix case,
the coefficients appear in pairs, leading to a bilinear system (BLS) of equations.
A summary of current approaches for solving BLS problems can be found in
\cite{Johnson:2014}.

First, let's count the number of data inputs and the number of free parameters
to see whether such a solution can exist.  Given $2 L$ Hermitian $\bm{C}(t)$,
there are $2 L N (N+1)/ 2 = L N (N+1)$ inputs, $L N$ eigenvalues $\lambda_k$
and $L N^2$ matrix elements $z_{n k}$.  So the equal number of inputs
and free parameters suggest a solution is possible.  If we naively construct
the block Vandermonde system analogous to Eq.~(\ref{eq:Vandermonde})
\begin{equation}
\label{eq:Vandermonde_block}
\left[\begin{array}{c}
  \bm{C}_0     \\
  \bm{C}_1     \\
  \vdots       \\
  \bm{C}_{2 L-1} \\
\end{array}\right] = \left[\begin{array}{ccc}
\bm{I}                 & \cdots & \bm{I}                 \\
\lambda_1 \bm{I}       & \cdots & \lambda_{N L} \bm{I}       \\
\vdots                 & \ddots & \vdots                 \\
\lambda_1^{2 L-1} \bm{I} & \cdots & \lambda_{L N}^{2 L-1} \bm{I} \\
\end{array}\right] \left[\begin{array}{c}
  \bm{A}_1  \\
  \vdots    \\
  \bm{A}_{L N}
\end{array}\right]
\end{equation}
the input data is $\mathbb{C}^{2 L N \times N}$,
the solution is $\mathbb{C}^{L N^2 \times N}$. So, a unique solution exists
by SVD if $N \le 2$ and for $N=2$ the $\bm{A}_k$ should be rank-1.
In our numerical tests we have found that the $N=2$ solution is always rank-1,
perhaps because our test function always started from a rank-1 construction,
but we don't have a proof that it should always be the case.
For $N \ge 3$, the system is underconstrained so an 
infinite number of general rank solutions exist but it is still possible
to find a unique rank-1 solution \cite{Johnson:2014}.

In our particular case, where the number of equations equals the number
of unknowns it should be possible to construct a solution where one unknown
is solved for each matrix element and the result used to eliminate that
unknown from the remaining system of equations by substitution.  For example,
start with
\begin{equation}
C_{12}(0) = \sum_{k=1}^{L N} z_{1k} z_{2k}^*
\end{equation}
and solve for $z_{11}$
\begin{equation}
z_{11} = \frac{1}{z_{21}^*} \left[
C_{12}(0) - \sum_{k=2}^{L N} z_{1 k} z_{2 k}^*
\right]
\end{equation}
Next, write the equation for $C_{13}(0)$, eliminate $z_{11}$ by substitution
\begin{equation}
C_{13}(0) = \frac{z_{31}^*}{z_{21}^*} \left[
C_{12}(0) - \sum_{k=2}^{L N} z_{1 k} z_{2 k}^*
\right] + \sum_{k^\prime=2}^{L N} z_{1 k^\prime} z_{3 k^\prime}
\end{equation}
and then solve for $z_{12}$, \textit{etc}.
The challenge of this approach is that the reduced
set of equations will become increasingly higher polynomial order
in the remaining unknowns.  Still, with a sufficiently small number number
of unknowns, it should be possible for symbolic math program like
\texttt{Mathematica} to find a solution.

Another possibility would be to increase the amount of input data
to $L N$ equally spaced $\bm{C}_t$ and form an over-constrained Hankel system
Eq.~(\ref{eq:linear_prediction_hankel_system_block}) to extract using SVD
a least-squares estimate of the same
$L$ $\bm{P}_\ell$ coefficient matrices.  Now, a square Vandermonde system
Eq.~(\ref{eq:Vandermonde_block}) can be formed for any $N$ and solved uniquely.
There is no guarantee that the $\bm{A}_k$ matrices found this way
are rank-1.  Also, the amount of input data exceeds the number
of free parameters of a rank-1 solution, so even if a rank-1 solution
is constructed it should be considered, at best, a least-squares solution.
Another consideration is that in Euclidean lattice field theory calculations
it is often computationally less expensive to increase $N$ than it is to increase
the number of available times $t$.  It is even worse if one considers
the signal-to-noise of such correlation matrices typically decreases
exponentially with increasing time separations
\cite{Parisi:1983ae, Lepage:1989hd}.

In the absence of an exact method, the method we will advocate is given
$\bm{C}_t$ and $\lambda_k$ following the block Prony method, use non-linear
least squares minimization to find the solution for the $z_{n k}$.
If the numerical residual is sufficiently close to zero, we will take that
as an indication that we have found the unique rank-1 solution.  Setting up
the minimization is not hard since the gradients and Hessian of the residual
with respect to $z_{n k}$ are easy to compute.

\section{\label{sec:error_free_examples}Error-Free Examples}

First we will consider an $N=2$ example of $L=6$ polynomial order.  We choose
the spectrum to be equally-spaced in the interval $0 < \lambda_k < 1$
\begin{equation}
\label{eq:spectrum}
\lambda_k = \frac{2 L + 1 - k}{2 L + 1} \in
\left\{ \frac{12}{13}, \frac{11}{13}, \cdots , \frac{1}{13} \right\} .
\end{equation}
The matrix elements $z_{n k}$ are drawn at random in \texttt{Mathematica}
using \texttt{SeedRandom[11]} and
\texttt{Transpose[RandomInteger[\{-9,9\},\{$L$ $N$,$N$\}]]}
\begin{widetext}
\begin{equation}
\bm{Z} = \left[
\begin{array}{rrrrrrrrrrrrrrrr}
 -3 & -1 &  6 & -3 & 6 & 4 &  4 & -6 & 6 &  0 & -6 & 6 \\
 -3 &  2 & -8 & -5 & 6 & 5 & -6 &  6 & 9 & -1 & -7 & 4 \\
\end{array}
\right]
\end{equation}
\end{widetext}
$\det \bm{Z} \bm{Z}^\dagger = 93894$ confirms the linear independence
of the rows.  With these inputs
\begin{equation}
\bm{C}_0 = \left[
\begin{array}{rr}
267 &  90 \\
 90 & 382
\end{array}
\right]
\end{equation}
\begin{equation}
\bm{C}_1 = \left[
\begin{array}{rr}
\frac{2763}{17} & \frac{437}{17} \\
\frac{437}{17} & \frac{4040}{17}
\end{array}
\right]
\end{equation}
and so forth. The reader can verify with these data that the block Prony method
reproduces the spectrum and the solution to Eq.~(\ref{eq:Vandermonde_block})
reproduces the correct rank-1 solution $\bm{Z}$.  As mentioned previously,
we do not have a proof for $N=2$ that the solution will be rank-1 but
it is the case in this example and all other random examples we tried
for various sizes $L$.

Next, we consider an $N=3$ example of $L=4$ polynomial order so we can use
the same input spectrum as Eq.~(\ref{eq:spectrum}). The matrix elements
$z_{n k}$ are drawn at random in \texttt{Mathematica}
using \texttt{SeedRandom[13]} and
\texttt{Transpose[RandomInteger[\{-9,9\},\{$L$ $N$,$N$\}]]}
\begin{widetext}
\begin{equation}
\bm{Z} = \left[
\begin{array}{rrrrrrrrrrrr}
 5 &  1 & 8 &  4 & 0 &  2 & -8 &  6 & -4 &  8 &  5 & -7 \\
 6 &  5 & 1 & -7 & 4 &  8 &  0 & -6 &  1 & -7 & -2 & -5 \\
 6 & -1 & 0 &  2 & 8 & -5 & -6 & -4 &  1 & -7 & -9 &  9 \\
\end{array}
\right]
\end{equation}
\end{widetext}
$\det \bm{Z} \bm{Z}^\dagger = 38,448,718$ confirms the linear independence
of the rows.  With these inputs
\begin{equation}
\bm{C}_0 = \left[
\begin{array}{rrr}
 364 & -40 & -117 \\
 -40 & 306 & 56 \\
 -117 & 56 & 394 \\
\end{array}
\right]
\end{equation}
\begin{equation}
\label{eq:ZN3L4}
\bm{C}_1 = \left[
\begin{array}{rrr}
 \frac{2042}{13} & \frac{6}{13} & 14 \\
 \frac{6}{13} & \frac{2098}{13} & \frac{489}{13} \\
 14 & \frac{489}{13} & \frac{1856}{13} \\
\end{array}
\right]
\end{equation}
and so forth.  Again, the reader can verify the block Prony method reproduces
the spectrum.  Eq.~(\ref{eq:Vandermonde_block}) cannot be used to compute
the amplitudes as it is under-determined.  If we define the components
of the residual vector from Eq.(\ref{eq:corr})
\begin{equation}
\label{eq:residual}
r_{abt} = 
C_{ab}(t) - \sum_{k=1}^{L N} z_{zk} z_{bk}^* \lambda_k^t
\end{equation}
Then minimizing the squared norm of the residual vector
\begin{equation}
r^2 = \sum_{a=1}^N \sum_{b \ge a} \sum_{t=0}^{2 L - 1} | r_{abt} |^2
\end{equation}
using \texttt{Mathematica}'s \texttt{NMinimize[]}
with 30 digits of precision finds $r \approx 0.00089$ and
\begin{widetext}
\begin{equation}
\bm{Z} \approx \left[
\begin{array}{rrrrrrrrrrrr}
5.0 & -0.94 & -8.0 & -4.2 & -0.096 & 1.8 & 7.6 & -7.1 & -2.3 & -8.1 & 5.2 & -7.0 \\
6.0 & -5.0 & -1.1 & 6.8 & 4.6 & 8.0 & 0.48 & 5.0 & 4.2 & 6.3 & -2.2 & -5.0 \\
6.0 & 0.95 & 0.12 & -2.3 & 7.8 & -5.3 & 6.1 & 3.0 & 3.2 & 6.5 & -9.2 & 9.0 \\
\end{array}
\right]
\end{equation}
\end{widetext}
whereas computing the residual using the computed eigenvalues and correct amplitudes
from Eq.~(\ref{eq:ZN3L4}) gives zero to machine precision.   
Clearly finding the global
minimum of the residual will be a challenging optimization problem.

\section{Variational Analysis}

In practical applications, the number of eigenstates exceeds the number of operators
that is typically constructed so any eigenvalues computed using the block Prony method
(BPM) 
will, at best, be variational estimates.  In Sec.~\ref{sub:GEM} we discussed various
bounds and asymptotic limits of generalized eigenvalues which confirm their variational
nature.  We would like to test the BPM eigenvalues in a similar
fashion.  We continue to consider an error-free correlation function constructed
from 12 states as in Eq.~(\ref{eq:spectrum}) and generate various combinations
of $L N = 12$.  We then assume the polynomial order is $K = L - 1$ and compute
$K N$ BPM eigenvalues for $t_\text{min} = t_\text{max} - 2 K + 1$,
$t \in [ t_\text{min} , t_\text{max} ]$.

\begin{figure}[ht]
    \centering
    \includegraphics[width=0.48\textwidth]{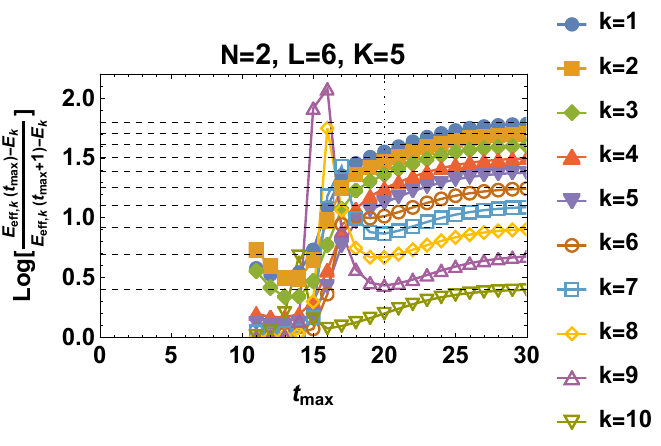}
    \caption{\label{fig:BPMN2L6K5err}
    Effective energies for $N=2, L=6$ example
    of Sec.~\ref{sec:error_free_examples}, assuming $K=5$ and using
    $t_\text{min} = t_\text{max} - 2 K + 1$, 
    $t \in [ t_\text{min} , t_\text{max} ]$.  Horizontal dashed lines
    are $E_{N K + 1} - E_k$ and points to the right of vertical dotted line
    satisfy $t_\text{min} > t_\text{max} / 2$ \cite{Blossier:2009kd}.
    }
\end{figure}

\begin{figure}[ht]
    \centering
    \includegraphics[width=0.48\textwidth]{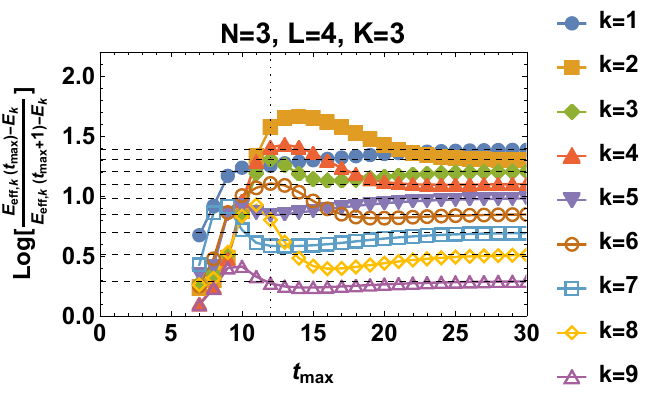}
    \caption{\label{fig:BPMN3L4K3err}
    Effective energies for $N=3, L=4$ example
    of Sec.~\ref{sec:error_free_examples}, assuming $K=3$ and using
    $t_\text{min} = t_\text{max} - 2 K + 1$, 
    $t \in [ t_\text{min} , t_\text{max} ]$.  Horizontal dashed lines
    are $E_{N K + 1} - E_k$ and points to the right of vertical dotted line
    satisfy $t_\text{min} > t_\text{max} / 2$ \cite{Blossier:2009kd}.
    }
\end{figure}

\begin{figure}[ht]
    \centering
    \includegraphics[width=0.48\textwidth]{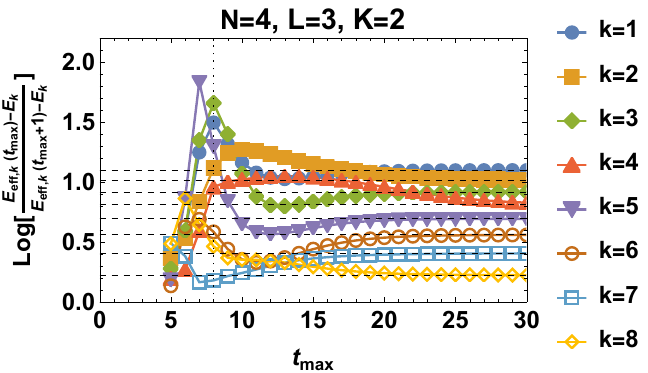}
    \caption{\label{fig:BPMN4L3K2err}
    Effective energies for $N=4, K=2$ using same example spectrum
    of Sec.~\ref{sec:error_free_examples} and using
    $t_\text{min} = t_\text{max} - 2 K + 1$, 
    $t \in [ t_\text{min} , t_\text{max} ]$.  Horizontal dashed lines
    are $E_{N K + 1} - E_k$ and points to the right of vertical dotted line
    satisfy $t_\text{min} > t_\text{max} / 2$ \cite{Blossier:2009kd}.
    }
\end{figure}

\begin{figure}[ht]
    \centering
    \includegraphics[width=0.48\textwidth]{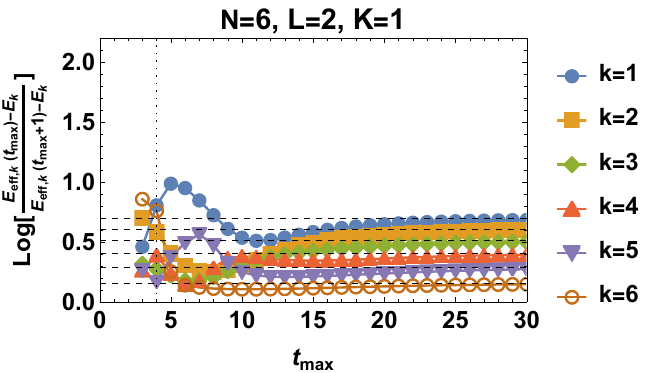}
    \caption{\label{fig:BPMN6L2K1err}
    Effective energies for $N=6, K=1$ using same example spectrum
    of Sec.~\ref{sec:error_free_examples} and using
    $t_\text{min} = t_\text{max} - 2 K + 1$, 
    $t \in [ t_\text{min} , t_\text{max} ]$.  Horizontal dashed lines
    are $E_{N K + 1} - E_k$ and points to the right of vertical dotted line
    satisfy $t_\text{min} > t_\text{max} / 2$ \cite{Blossier:2009kd}.
    }
\end{figure}

For any time range, the eigenvalues are
always bounded by the interlacing theorem.  They also approach the true values
as $t_\text{max} \to \infty$.  In Figs.~\ref{fig:BPMN2L6K5err}, \ref{fig:BPMN3L4K3err},
\ref{fig:BPMN4L3K2err} and \ref{fig:BPMN6L2K1err} we compute the exponential decay
rate of the error of the effective energies and compare with the asymptotic estimate
in Eq.~(\ref{eq:eff_energy_error}).  In all cases, the decay rate of the error
approaches the estimate.  In the figures we plot a vertical line where
points to the right satisfy
$t_\text{min} > t_\text{max} / 2$. In \cite{Blossier:2009kd}, this was considered
to be a necessary condition for the asymptotic decay rate to dominate.  In the figures,
it seems that this is not always sufficient, particularly if it occurs at small time
values.
From this study, we conclude that the BPM eigenvalues are natural extensions
of the generalized eigenvalues to higher polynomial order and are variational estimates
of the true eigenvalues with well-defined asymptotic behavior.

\section{\label{sec:noisy_data}Comments on Modeling Noisy Data}

In practical applications, we must consider how best to apply the block Prony
method (BPM) to model noisy data.  In previous studies of this problem
\cite{Fleming:2009wb, Cushman:2019hfh}, because the algebraic method maps all the
data into a complete set of model parameters precisely describing the data,
we focused on distinguishing which model parameters encoded mostly the signal
present in the data and which encoded mostly the noise.

Recently, the method of \textit{Bayesian model averaging} has been developed
\cite{Jay:2020jkz, Neil:2022joj, Neil:2023pgt} for analysis of lattice field
theory results and provides us with a tool for weighing how much information
is represented by a given subset of model parameters.  For this discussion,
we will define the model probability
\begin{equation}
p(M|D) \propto \exp( - \chi^2 - 2 q )
\end{equation}
and the $\chi^2$ statistic is computed from residuals of Eq.~(\ref{eq:residual})
and some estimate of the inverse of the data covariance matrix in the usual way.
In the sense described in \cite{Neil:2022joj}, the full set of BPM parameters
describe a ``perfect'' model in that $\chi^2 = 0$.  Even though the BPM model
describes the data perfectly, it is unlikely maximize $p(M|D)$. To see this,
imagine a state $k$ encodes mostly noise.  Removing parameters $\lambda_k$
and $z_{nk}$ will reduce $2 q$ by $2(N+1)$ but will not increase $\chi^2$ by
anywhere near the same amount due to the noise.  Thus, the model with state $k$
removed will be more probable that the full model.  So, our heuristic 
in dealing with noisy data will be to partition the BPM parameters by states
which are mostly signal and mostly noise by finding the subset of states $k$
which maximizes the model probability.  We will describe this process more
completely in a future work.

\section{\label{sec:conclusion}Conclusion}

In this work, we have noted that the generalized eigenvalue method (GEM)
is an extension of the usual effective energy method to Hermitian matrix-valued
correlation functions and can be reformulated as a first-order matrix polynomial.
We then extend GEM to higher-order matrix polynomials by finding the generalization
of Prony's method to Hermitian matrix-valued correlation functions, which we call
the block Prony method (BPM).  The companion matrix plays a central role as it
enables us to understand the space of solutions of the matrix polynomial
as the eigenspace of the companion matrix.
We provided some error-free numerical examples, including a demonstration that
the BPM eigenvalues are variational estimates with the expected
asymptotic error bounds \cite{Blossier:2009kd}.  We also provided some comments
on our future plans for applying the BPM to the analysis of data with noise.

\bibliography{main}

\providecommand{\noopsort}[1]{}\providecommand{\singleletter}[1]{#1}%
\begin{thebibliography}{20}%
\makeatletter
\providecommand \@ifxundefined [1]{%
 \@ifx{#1\undefined}
}%
\providecommand \@ifnum [1]{%
 \ifnum #1\expandafter \@firstoftwo
 \else \expandafter \@secondoftwo
 \fi
}%
\providecommand \@ifx [1]{%
 \ifx #1\expandafter \@firstoftwo
 \else \expandafter \@secondoftwo
 \fi
}%
\providecommand \natexlab [1]{#1}%
\providecommand \enquote  [1]{``#1''}%
\providecommand \bibnamefont  [1]{#1}%
\providecommand \bibfnamefont [1]{#1}%
\providecommand \citenamefont [1]{#1}%
\providecommand \href@noop [0]{\@secondoftwo}%
\providecommand \href [0]{\begingroup \@sanitize@url \@href}%
\providecommand \@href[1]{\@@startlink{#1}\@@href}%
\providecommand \@@href[1]{\endgroup#1\@@endlink}%
\providecommand \@sanitize@url [0]{\catcode `\\12\catcode `\$12\catcode
  `\&12\catcode `\#12\catcode `\^12\catcode `\_12\catcode `\%12\relax}%
\providecommand \@@startlink[1]{}%
\providecommand \@@endlink[0]{}%
\providecommand \url  [0]{\begingroup\@sanitize@url \@url }%
\providecommand \@url [1]{\endgroup\@href {#1}{\urlprefix }}%
\providecommand \urlprefix  [0]{URL }%
\providecommand \Eprint [0]{\href }%
\providecommand \doibase [0]{https://doi.org/}%
\providecommand \selectlanguage [0]{\@gobble}%
\providecommand \bibinfo  [0]{\@secondoftwo}%
\providecommand \bibfield  [0]{\@secondoftwo}%
\providecommand \translation [1]{[#1]}%
\providecommand \BibitemOpen [0]{}%
\providecommand \bibitemStop [0]{}%
\providecommand \bibitemNoStop [0]{.\EOS\space}%
\providecommand \EOS [0]{\spacefactor3000\relax}%
\providecommand \BibitemShut  [1]{\csname bibitem#1\endcsname}%
\let\auto@bib@innerbib\@empty
\bibitem [{\citenamefont {Cohen}\ \emph {et~al.}(2010)\citenamefont {Cohen},
  \citenamefont {Fleming},\ and\ \citenamefont {Lin}}]{Cohen:2010}%
  \BibitemOpen
  \bibfield  {author} {\bibinfo {author} {\bibfnamefont {S.~D.}\ \bibnamefont
  {Cohen}}, \bibinfo {author} {\bibfnamefont {G.~T.}\ \bibnamefont {Fleming}},\
  and\ \bibinfo {author} {\bibfnamefont {H.-W.}\ \bibnamefont {Lin}},\
  }\bibinfo {title} {Exponential time series in lattice quantum field theory},\
  in\ \href {https://doi.org/10.2174/97816080504821100101} {\emph {\bibinfo
  {booktitle} {Exponential Data Fitting and Its Applications}}},\ \bibinfo
  {editor} {edited by\ \bibinfo {editor} {\bibfnamefont {V.}~\bibnamefont
  {Pereyra}}\ and\ \bibinfo {editor} {\bibfnamefont {G.}~\bibnamefont
  {Scherer}}}\ (\bibinfo  {publisher} {Bentham Science Publishers},\ \bibinfo
  {year} {2010})\ Chap.~\bibinfo {chapter} {4}, pp.\ \bibinfo {pages}
  {71--93}\BibitemShut {NoStop}%
\bibitem [{\citenamefont {Jay}\ and\ \citenamefont {Neil}(2021)}]{Jay:2020jkz}%
  \BibitemOpen
  \bibfield  {author} {\bibinfo {author} {\bibfnamefont {W.~I.}\ \bibnamefont
  {Jay}}\ and\ \bibinfo {author} {\bibfnamefont {E.~T.}\ \bibnamefont {Neil}},\
  }\bibfield  {title} {\bibinfo {title} {{Bayesian model averaging for analysis
  of lattice field theory results}},\ }\href
  {https://doi.org/10.1103/PhysRevD.103.114502} {\bibfield  {journal} {\bibinfo
   {journal} {Phys. Rev. D}\ }\textbf {\bibinfo {volume} {103}},\ \bibinfo
  {pages} {114502} (\bibinfo {year} {2021})},\ \Eprint
  {https://arxiv.org/abs/2008.01069} {arXiv:2008.01069 [stat.ME]} \BibitemShut
  {NoStop}%
\bibitem [{\citenamefont {Wilson}(1981)}]{Wilson:1981}%
  \BibitemOpen
  \bibfield  {author} {\bibinfo {author} {\bibfnamefont {K.~G.}\ \bibnamefont
  {Wilson}},\ }\bibfield  {title} {\bibinfo {title} {Closing remarks}}
  (\bibinfo {year} {1981}),\ \bibinfo {note} {{Abingdon Meeting on Lattice
  Gauge Theories}}\BibitemShut {NoStop}%
\bibitem [{\citenamefont {Berg}(1982)}]{Berg:1982hf}%
  \BibitemOpen
  \bibfield  {author} {\bibinfo {author} {\bibfnamefont {B.}~\bibnamefont
  {Berg}},\ }\bibfield  {title} {\bibinfo {title} {Glueball calculations in
  lattice gauge theories},\ }\href {https://doi.org/10.1051/jphyscol:1982355}
  {\bibfield  {journal} {\bibinfo  {journal} {J. Phys. Colloq.}\ }\textbf
  {\bibinfo {volume} {43}},\ \bibinfo {pages} {272} (\bibinfo {year}
  {1982})}\BibitemShut {NoStop}%
\bibitem [{\citenamefont {Michael}\ and\ \citenamefont
  {Teasdale}(1983)}]{Michael:1982gb}%
  \BibitemOpen
  \bibfield  {author} {\bibinfo {author} {\bibfnamefont {C.}~\bibnamefont
  {Michael}}\ and\ \bibinfo {author} {\bibfnamefont {I.}~\bibnamefont
  {Teasdale}},\ }\bibfield  {title} {\bibinfo {title} {{Extracting Glueball
  Masses From Lattice {QCD}}},\ }\href
  {https://doi.org/10.1016/0550-3213(83)90674-0} {\bibfield  {journal}
  {\bibinfo  {journal} {Nucl. Phys. B}\ }\textbf {\bibinfo {volume} {215}},\
  \bibinfo {pages} {433} (\bibinfo {year} {1983})}\BibitemShut {NoStop}%
\bibitem [{\citenamefont {Luscher}\ and\ \citenamefont
  {Wolff}(1990)}]{Luscher:1990ck}%
  \BibitemOpen
  \bibfield  {author} {\bibinfo {author} {\bibfnamefont {M.}~\bibnamefont
  {Luscher}}\ and\ \bibinfo {author} {\bibfnamefont {U.}~\bibnamefont
  {Wolff}},\ }\bibfield  {title} {\bibinfo {title} {{How to Calculate the
  Elastic Scattering Matrix in Two-dimensional Quantum Field Theories by
  Numerical Simulation}},\ }\href
  {https://doi.org/10.1016/0550-3213(90)90540-T} {\bibfield  {journal}
  {\bibinfo  {journal} {Nucl. Phys. B}\ }\textbf {\bibinfo {volume} {339}},\
  \bibinfo {pages} {222} (\bibinfo {year} {1990})}\BibitemShut {NoStop}%
\bibitem [{\citenamefont {Blossier}\ \emph {et~al.}(2009)\citenamefont
  {Blossier}, \citenamefont {Della~Morte}, \citenamefont {von Hippel},
  \citenamefont {Mendes},\ and\ \citenamefont {Sommer}}]{Blossier:2009kd}%
  \BibitemOpen
  \bibfield  {author} {\bibinfo {author} {\bibfnamefont {B.}~\bibnamefont
  {Blossier}}, \bibinfo {author} {\bibfnamefont {M.}~\bibnamefont
  {Della~Morte}}, \bibinfo {author} {\bibfnamefont {G.}~\bibnamefont {von
  Hippel}}, \bibinfo {author} {\bibfnamefont {T.}~\bibnamefont {Mendes}},\ and\
  \bibinfo {author} {\bibfnamefont {R.}~\bibnamefont {Sommer}},\ }\bibfield
  {title} {\bibinfo {title} {{On the generalized eigenvalue method for energies
  and matrix elements in lattice field theory}},\ }\href
  {https://doi.org/10.1088/1126-6708/2009/04/094} {\bibfield  {journal}
  {\bibinfo  {journal} {JHEP}\ }\textbf {\bibinfo {volume} {04}},\ \bibinfo
  {pages} {094}},\ \Eprint {https://arxiv.org/abs/0902.1265} {arXiv:0902.1265
  [hep-lat]} \BibitemShut {NoStop}%
\bibitem [{\citenamefont {Gohberg}\ \emph {et~al.}(2009)\citenamefont
  {Gohberg}, \citenamefont {Lancaster},\ and\ \citenamefont
  {Rodman}}]{Gohberg:2009}%
  \BibitemOpen
  \bibfield  {author} {\bibinfo {author} {\bibfnamefont {I.}~\bibnamefont
  {Gohberg}}, \bibinfo {author} {\bibfnamefont {P.}~\bibnamefont {Lancaster}},\
  and\ \bibinfo {author} {\bibfnamefont {L.}~\bibnamefont {Rodman}},\ }\href
  {https://doi.org/10.1137/1.9780898719024} {\emph {\bibinfo {title} {Matrix
  Polynomials}}},\ Classics in Applied Mathematics\ (\bibinfo  {publisher}
  {Society for Industrial and Applied Mathematics},\ \bibinfo {address}
  {Philadelphia},\ \bibinfo {year} {2009})\BibitemShut {NoStop}%
\bibitem [{\citenamefont {Fleming}(2005)}]{Fleming:2004hs}%
  \BibitemOpen
  \bibfield  {author} {\bibinfo {author} {\bibfnamefont {G.~T.}\ \bibnamefont
  {Fleming}},\ }\bibfield  {title} {\bibinfo {title} {What can lattice {QCD}
  theorists learn from {NMR} spectroscopists?},\ }in\ \href
  {https://doi.org/10.1007/3-540-28504-0} {\emph {\bibinfo {booktitle} {QCD and
  Numerical Analysis III}}},\ \bibinfo {series and number} {\bibinfo {series}
  {Lecture Notes in Computational Science and Engineering}\ No.~\bibinfo
  {number} {47}},\ \bibinfo {editor} {edited by\ \bibinfo {editor}
  {\bibfnamefont {A.}~\bibnamefont {Bori{\c{c}}i}}, \bibinfo {editor}
  {\bibfnamefont {A.}~\bibnamefont {Frommer}}, \bibinfo {editor} {\bibfnamefont
  {B.}~\bibnamefont {Jo{\'o}}}, \bibinfo {editor} {\bibfnamefont
  {A.}~\bibnamefont {Kennedy}},\ and\ \bibinfo {editor} {\bibfnamefont
  {B.}~\bibnamefont {Pendleton}}}\ (\bibinfo  {publisher} {Springer-Verlag},\
  \bibinfo {year} {2005})\ pp.\ \bibinfo {pages} {143--152},\ \Eprint
  {https://arxiv.org/abs/hep-lat/0403023} {hep-lat/0403023} \BibitemShut
  {NoStop}%
\bibitem [{\citenamefont {de~Prony}(1795)}]{Prony:1795}%
  \BibitemOpen
  \bibfield  {author} {\bibinfo {author} {\bibfnamefont {G.~C. F. M.~R.}\
  \bibnamefont {de~Prony}},\ }\bibfield  {title} {\bibinfo {title} {Essai
  exp{\'e}rimental et analytique sur les lois de la dilatabilit{\'e} et sur
  celles de la force expansive de la vapeur de l'eau et de la vapeur de
  l'alkool, {\`a} diff{\'e}rentes temp{\'e}ratures},\ }\href@noop {} {\bibfield
   {journal} {\bibinfo  {journal} {J. Ecole Poly.}\ }\textbf {\bibinfo {volume}
  {1}},\ \bibinfo {pages} {24} (\bibinfo {year} {1795})},\ \bibinfo {note}
  {partial translation available in {\cite{Vandevoorde:1996}}}\BibitemShut
  {NoStop}%
\bibitem [{\citenamefont {Fleming}\ \emph {et~al.}(2009)\citenamefont
  {Fleming}, \citenamefont {Cohen}, \citenamefont {Lin},\ and\ \citenamefont
  {Pereyra}}]{Fleming:2009wb}%
  \BibitemOpen
  \bibfield  {author} {\bibinfo {author} {\bibfnamefont {G.~T.}\ \bibnamefont
  {Fleming}}, \bibinfo {author} {\bibfnamefont {S.~D.}\ \bibnamefont {Cohen}},
  \bibinfo {author} {\bibfnamefont {H.-W.}\ \bibnamefont {Lin}},\ and\ \bibinfo
  {author} {\bibfnamefont {V.}~\bibnamefont {Pereyra}},\ }\bibfield  {title}
  {\bibinfo {title} {{Excited-State Effective Masses in Lattice QCD}},\ }\href
  {https://doi.org/10.1103/PhysRevD.80.074506} {\bibfield  {journal} {\bibinfo
  {journal} {Phys. Rev. D}\ }\textbf {\bibinfo {volume} {80}},\ \bibinfo
  {pages} {074506} (\bibinfo {year} {2009})},\ \Eprint
  {https://arxiv.org/abs/0903.2314} {arXiv:0903.2314 [hep-lat]} \BibitemShut
  {NoStop}%
\bibitem [{\citenamefont {Beane}\ \emph {et~al.}(2009)\citenamefont {Beane},
  \citenamefont {Detmold}, \citenamefont {Luu}, \citenamefont {Orginos},
  \citenamefont {Parreno}, \citenamefont {Savage}, \citenamefont {Torok},\ and\
  \citenamefont {Walker-Loud}}]{Beane:2009kya}%
  \BibitemOpen
  \bibfield  {author} {\bibinfo {author} {\bibfnamefont {S.~R.}\ \bibnamefont
  {Beane}}, \bibinfo {author} {\bibfnamefont {W.}~\bibnamefont {Detmold}},
  \bibinfo {author} {\bibfnamefont {T.~C.}\ \bibnamefont {Luu}}, \bibinfo
  {author} {\bibfnamefont {K.}~\bibnamefont {Orginos}}, \bibinfo {author}
  {\bibfnamefont {A.}~\bibnamefont {Parreno}}, \bibinfo {author} {\bibfnamefont
  {M.~J.}\ \bibnamefont {Savage}}, \bibinfo {author} {\bibfnamefont
  {A.}~\bibnamefont {Torok}},\ and\ \bibinfo {author} {\bibfnamefont
  {A.}~\bibnamefont {Walker-Loud}},\ }\bibfield  {title} {\bibinfo {title}
  {{High Statistics Analysis using Anisotropic Clover Lattices: (I) Single
  Hadron Correlation Functions}},\ }\href
  {https://doi.org/10.1103/PhysRevD.79.114502} {\bibfield  {journal} {\bibinfo
  {journal} {Phys. Rev. D}\ }\textbf {\bibinfo {volume} {79}},\ \bibinfo
  {pages} {114502} (\bibinfo {year} {2009})},\ \Eprint
  {https://arxiv.org/abs/0903.2990} {arXiv:0903.2990 [hep-lat]} \BibitemShut
  {NoStop}%
\bibitem [{\citenamefont {Cushman}\ and\ \citenamefont
  {Fleming}(2020)}]{Cushman:2019hfh}%
  \BibitemOpen
  \bibfield  {author} {\bibinfo {author} {\bibfnamefont {K.~K.}\ \bibnamefont
  {Cushman}}\ and\ \bibinfo {author} {\bibfnamefont {G.~T.}\ \bibnamefont
  {Fleming}},\ }\bibfield  {title} {\bibinfo {title} {{Automated label flows
  for excited states of correlation functions in lattice gauge theory}},\
  }\href {https://doi.org/10.1103/PhysRevE.102.043303} {\bibfield  {journal}
  {\bibinfo  {journal} {Phys. Rev. E}\ }\textbf {\bibinfo {volume} {102}},\
  \bibinfo {pages} {043303} (\bibinfo {year} {2020})},\ \Eprint
  {https://arxiv.org/abs/1912.08205} {arXiv:1912.08205 [hep-lat]} \BibitemShut
  {NoStop}%
\bibitem [{\citenamefont {Fischer}\ \emph {et~al.}(2020)\citenamefont
  {Fischer}, \citenamefont {Kostrzewa}, \citenamefont {Ostmeyer}, \citenamefont
  {Ottnad}, \citenamefont {Ueding},\ and\ \citenamefont
  {Urbach}}]{Fischer:2020bgv}%
  \BibitemOpen
  \bibfield  {author} {\bibinfo {author} {\bibfnamefont {M.}~\bibnamefont
  {Fischer}}, \bibinfo {author} {\bibfnamefont {B.}~\bibnamefont {Kostrzewa}},
  \bibinfo {author} {\bibfnamefont {J.}~\bibnamefont {Ostmeyer}}, \bibinfo
  {author} {\bibfnamefont {K.}~\bibnamefont {Ottnad}}, \bibinfo {author}
  {\bibfnamefont {M.}~\bibnamefont {Ueding}},\ and\ \bibinfo {author}
  {\bibfnamefont {C.}~\bibnamefont {Urbach}},\ }\bibfield  {title} {\bibinfo
  {title} {{On the generalised eigenvalue method and its relation to Prony and
  generalised pencil of function methods}},\ }\href
  {https://doi.org/10.1140/epja/s10050-020-00205-w} {\bibfield  {journal}
  {\bibinfo  {journal} {Eur. Phys. J. A}\ }\textbf {\bibinfo {volume} {56}},\
  \bibinfo {pages} {206} (\bibinfo {year} {2020})},\ \Eprint
  {https://arxiv.org/abs/2004.10472} {arXiv:2004.10472 [hep-lat]} \BibitemShut
  {NoStop}%
\bibitem [{\citenamefont {Johnson}\ \emph {et~al.}(2014)\citenamefont
  {Johnson}, \citenamefont {{\v{S}}migoc},\ and\ \citenamefont
  {Yang}}]{Johnson:2014}%
  \BibitemOpen
  \bibfield  {author} {\bibinfo {author} {\bibfnamefont {C.~R.}\ \bibnamefont
  {Johnson}}, \bibinfo {author} {\bibfnamefont {H.}~\bibnamefont
  {{\v{S}}migoc}},\ and\ \bibinfo {author} {\bibfnamefont {D.}~\bibnamefont
  {Yang}},\ }\bibfield  {title} {\bibinfo {title} {Solution theory for systems
  of bilinear equations},\ }\href
  {https://doi.org/10.1080/03081087.2013.839670} {\bibfield  {journal}
  {\bibinfo  {journal} {Linear and Multilinear Algebra}\ }\textbf {\bibinfo
  {volume} {62}},\ \bibinfo {pages} {1553} (\bibinfo {year} {2014})},\ \Eprint
  {https://arxiv.org/abs/1303.4988} {arXiv:1303.4988 [math.RA]} \BibitemShut
  {NoStop}%
\bibitem [{\citenamefont {Parisi}(1984)}]{Parisi:1983ae}%
  \BibitemOpen
  \bibfield  {author} {\bibinfo {author} {\bibfnamefont {G.}~\bibnamefont
  {Parisi}},\ }\bibfield  {title} {\bibinfo {title} {{The Strategy for
  Computing the Hadronic Mass Spectrum}},\ }\href
  {https://doi.org/10.1016/0370-1573(84)90081-4} {\bibfield  {journal}
  {\bibinfo  {journal} {Phys. Rept.}\ }\textbf {\bibinfo {volume} {103}},\
  \bibinfo {pages} {203} (\bibinfo {year} {1984})}\BibitemShut {NoStop}%
\bibitem [{\citenamefont {Lepage}(1989)}]{Lepage:1989hd}%
  \BibitemOpen
  \bibfield  {author} {\bibinfo {author} {\bibfnamefont {G.~P.}\ \bibnamefont
  {Lepage}},\ }\bibfield  {title} {\bibinfo {title} {{The Analysis of
  Algorithms for Lattice Field Theory}},\ }in\ \href@noop {} {\emph {\bibinfo
  {booktitle} {{From Actions to Answers}}}},\ \bibinfo {series and number}
  {{Proceedings of the Theoretical Advanced Study Institute in Elementary
  Particle Physics}}\ (\bibinfo  {publisher} {{World Scientific}},\ \bibinfo
  {year} {1989})\ \bibinfo {note}
  {\url{https://lib-extopc.kek.jp/preprints/PDF/1990/9003/9003479.pdf}}\BibitemShut
  {NoStop}%
\bibitem [{\citenamefont {Neil}\ and\ \citenamefont
  {Sitison}(2022)}]{Neil:2022joj}%
  \BibitemOpen
  \bibfield  {author} {\bibinfo {author} {\bibfnamefont {E.~T.}\ \bibnamefont
  {Neil}}\ and\ \bibinfo {author} {\bibfnamefont {J.~W.}\ \bibnamefont
  {Sitison}},\ }\bibfield  {title} {\bibinfo {title} {{Improved information
  criteria for Bayesian model averaging in lattice field theory}},\ }\href@noop
  {} {\  (\bibinfo {year} {2022})},\ \Eprint {https://arxiv.org/abs/2208.14983}
  {arXiv:2208.14983 [stat.ME]} \BibitemShut {NoStop}%
\bibitem [{\citenamefont {Neil}\ and\ \citenamefont
  {Sitison}(2023)}]{Neil:2023pgt}%
  \BibitemOpen
  \bibfield  {author} {\bibinfo {author} {\bibfnamefont {E.~T.}\ \bibnamefont
  {Neil}}\ and\ \bibinfo {author} {\bibfnamefont {J.~W.}\ \bibnamefont
  {Sitison}},\ }\bibfield  {title} {\bibinfo {title} {{Model averaging
  approaches to data subset selection}},\ }\href@noop {} {\  (\bibinfo {year}
  {2023})},\ \Eprint {https://arxiv.org/abs/2305.19417} {arXiv:2305.19417
  [stat.ME]} \BibitemShut {NoStop}%
\bibitem [{\citenamefont {Vandevoorde}(1996)}]{Vandevoorde:1996}%
  \BibitemOpen
  \bibfield  {author} {\bibinfo {author} {\bibfnamefont {D.}~\bibnamefont
  {Vandevoorde}},\ }\emph {\bibinfo {title} {A fast exponential decomposition
  algorithm and its applications to structured matrices}},\ \href@noop {}
  {Ph.D. thesis},\ \bibinfo  {school} {Rensselaer Polytechnic Institute}
  (\bibinfo {year} {1996}),\ \bibinfo {note}
  {{\url{http://wwwlib.umi.com/dissertations/fullcit/9806112}}}\BibitemShut
  {NoStop}%
\end{thebibliography}%

\end{document}